\documentclass[fleqn,10pt]{wlscirep}

\usepackage{color,soul}
\usepackage{upgreek}
\usepackage{textgreek}

\title{Reprogrammable Graphene-based Metasurface Mirror with Adaptive Focal Point for THz Imaging}

\author[1,2*]{Seyed Ehsan Hosseininejad}
\author[3]{Kasra Rouhi}
\author[2]{Mohammad Neshat}
\author[2]{Reza Faraji-Dana}
\author[4]{Albert Cabellos-Aparicio}
\author[4]{Sergi Abadal}
\author[4]{Eduard Alarc\'{o}n}
\affil[1]{Department of Electrical  Engineering,  Yazd  University,  Yazd, Iran}
\affil[2]{School of Electrical and Computer Engineering, University of Tehran, Tehran, Iran}
\affil[3]{School of Electrical Engineering, Iran University of Science and Technology, Tehran, Iran}
\affil[4]{NaNoNetworking Center in Catalonia (N3Cat), Universitat Polit\`{e}cnica de Catalunya, 08034 Barcelona, Spain}

\affil[*]{sehosseininejad@yazd.ac.ir; sehosseininejad@ut.ac.ir}


\begin{abstract}
Recent emergence of metasurfaces has enabled the development of ultra-thin flat optical components through different wavefront shaping techniques at various wavelengths. 
However, due to the non-adaptive nature of conventional metasurfaces, the focal point of the resulting optics needs to be fixed at the design stage, thus severely limiting its reconfigurability and applicability. In this paper, we aim to overcome such constraint by presenting a flat reflective component that can be reprogrammed to focus terahertz waves at a desired point in the near-field region. To this end, we first propose a graphene-based unit cell with phase reconfigurability, and then employ the coding metasurface approach to draw the phase profile required to set the focus on the target point. Our results show that the proposed component can operate close to the diffraction limit with high focusing range and low focusing error. We also demonstrate that, through appropriate automation, the reprogrammability of the metamirror could be leveraged to develop compact terahertz scanning and imaging systems, as well as novel reconfigurable components for terahertz wireless communications.
\end{abstract}
\begin{document}

\flushbottom
\maketitle
%
%
\thispagestyle{empty}


\section*{Introduction}
Metamaterials have drawn a great deal of attention since their conception as they enable unprecedented levels of electromagnetic control \cite{Engheta2006}. They have led to significant breakthroughs in various fields such as imaging, radar, and wireless communications to name a few \cite{Glybovski2016, Chen2016, Vellucci2017}. Metasurfaces, the thin-film planar analogue of metamaterials, are composed of an array of subwavelength resonators, and inherit the unique properties of their three-dimensional counterparts while addressing their issues related to bulkiness, losses, and cost. For this, metasurfaces operating at the microwave \cite{Yang2016, Wan2016, Li2017b, Tcvetkova2018}, Terahertz \cite{Qu2015, Zhang2016a, Liu2016a, Qu2017}, or optical \cite{ChenXZ2012, Li2015, Glybovski2016} bands have been widely proposed to achieve attractive features such as negative refraction, cloaking, superlensing, or holographic behavior. 
A large subset of these designs have been based on the generalization of the Snell's laws of reflection and refraction \cite{Yu2011a}, which provided fundamental understanding on how to achieve precise beam manipulation \cite{Huang2017}, vortex creation \cite{Shi2017}, focusing \cite{Yang2016} through the drawing of specific phase gradients. Another remarkable advance is the proposal of coding metasurfaces, where the device is built (\emph{encoded}) using a discrete set of unit cell options (\emph{bits} or \emph{words}) \cite{Cui2014, liu2017concepts}. This provides a powerful and intuitive design perspective while drawing a clear parallelism with information theory, which opens new ways to model, compose, and design advanced metasurfaces \cite{Cui2016, Liu2016b}. Several works have exemplified such approach mainly for scattering control \cite{Gao2015, Liu2016a, Zhang2016a}.

Two of the main downturns of most metasurfaces are non-adaptivity and non-reconfigurability as, in most cases, the electromagnetic function and its scope are fixed once the unit cell is designed. 
In response to these drawbacks, metasurface designs with tunable or switchable elements in the unit cell have emerged \cite{Oliveri2015}. The resulting reconfigurable metasurfaces can be globally or locally tunable, depending on the specific scheme and, with appropriate control means, they become programmable \cite{Liu2018ISCAS}.
Not coincidentally, the programmable metasurface paradigm has found in the aforementioned coding paradigm a natural match to describe and implement reconfigurability. When built using locally switchable elements such as pin diodes, varactors, or switches, coding metasurfaces can be elegantly described as a bit or state matrix and digitally controlled through reconfigurable logic, e.g. a Field-Programmable Gate Array (FPGA) \cite{Cui2014, Wan2016}. Polarization and focusing control \cite{Yang2016}, beam manipulation \cite{Wan2016, Huang2017}, or holograms \cite{Li2017b} have been demonstrated in the GHz range, whereas fewer proposals have appeared in the THz regime mostly due to the lack of appropriate means for tuning.

Graphene shows great promise for the implementation of THz metasurfaces. This is due to its widely demonstrated plasmonic properties that provide subwavelength behavior required in the unit cell building block \cite{hosseininejad2016comparative,hosseininejad2017study}. The reasonably low loss of graphene plasmons at THz frequency band has been the main motivation of several wave steering devices \cite{Li2015, Deng2017}. The plasmon tunability of graphene is achievable through chemical doping or electrostatic biasing. The tunability of graphene opens the door to globally reconfigurable metasurfaces with fine-grained control of the amplitude-phase response \cite{Carrasco2013a}, or it can provide switching between diverse electromagnetic functionalities in particular setups \cite{Biswas2018}. Plasmon tunability can also be leveraged locally, in which case graphene becomes a natural candidate for the implementation of (re)programmable THz metasurfaces. In spite of their potential, graphene-based programmable designs have remained relatively unexplored, and current proposals are mainly limited to dynamic beam manipulation for diffusion \cite{Rouhi2017a}, encryption \cite{Momeni2018}, or vorticity control \cite{Shi2017}.

\begin{figure}[!ht] 
\centering
\includegraphics[width=1\columnwidth]{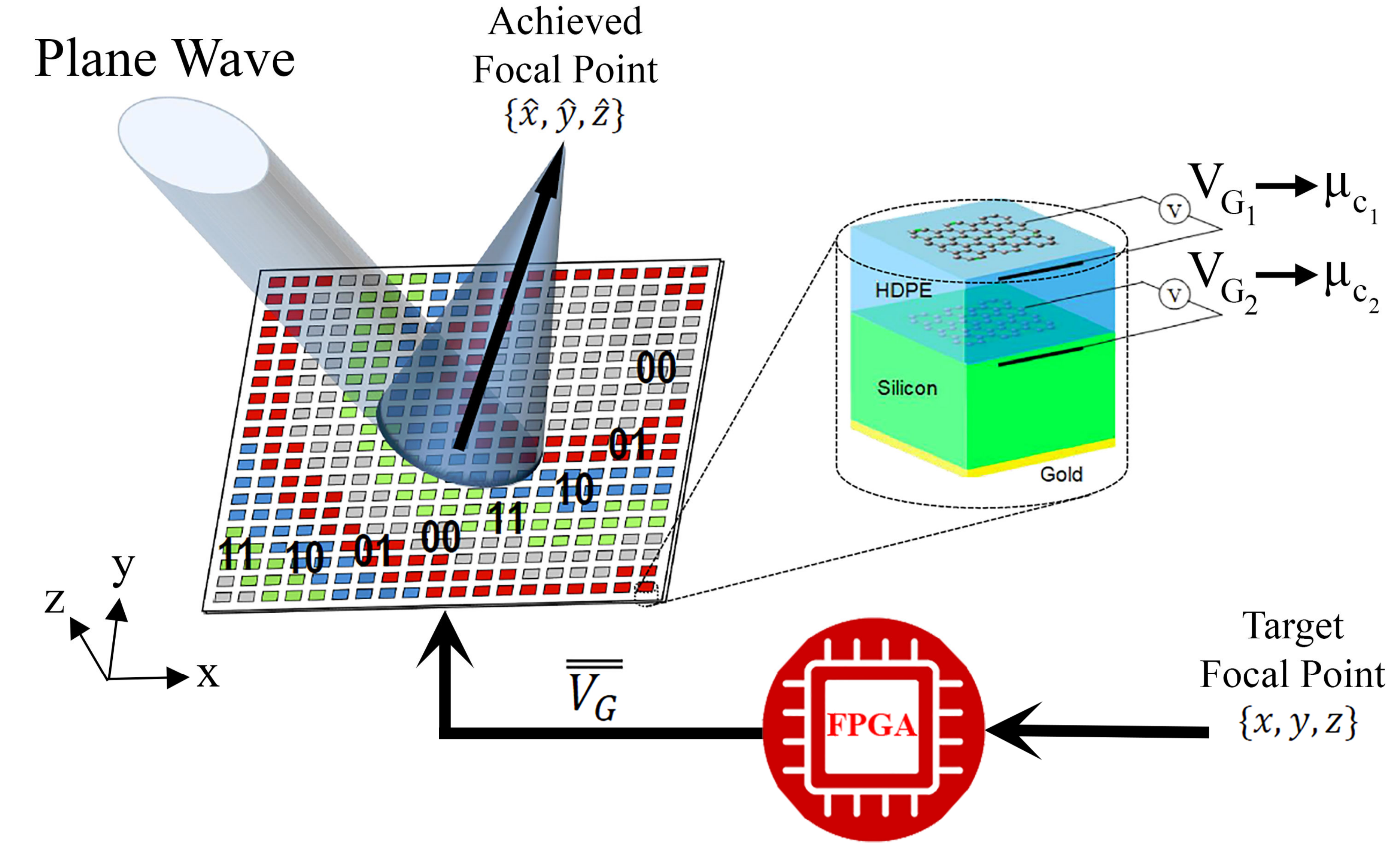} 
\vspace{-0.1cm}
\caption{Flat mirror coding metasurface with tunable focal point. The desired focal point $\{x,y,z\}$ is translated into a state matrix by a programmable device such as an FPGA. The state matrix is a digital representation of the phase profile of the metasurface, and is translated into voltage matrix $\overline{\overline{V_{G}}}$ with a discrete set of bias voltages. The bias voltages drive the unit cells, to achieve a 360\textsuperscript{o} phase range thanks to its dual layer structure and the tuning of their chemical potentials $\mu_{c1}$, $\mu_{c2}$. When the metasurface is illuminated by a plane wave, the energy is focused on the target focal point $\{\hat{x},\hat{y},\hat{z}\}$.}
\label{fig:metasurface}
\vspace{-0.5cm}
\end{figure}

In this work, a (re)programmable metamirror is proposed as shown in Figure \ref{fig:metasurface}. The device is conceived as a 2-bit coding metasurface that leverages the tunability of its graphene-based unit cells to control both the position and depth of focus. The shift of focus is achieved through a modification of the metasurface phase profile, that is achieved by simply changing the bias applied to individual unit cells. The generalized Snell's law of reflection allows to calculate the exact phase profile required for each target focal point, thus opening the door to fast precise focusing. If the change in the phase profile is automatized, it can be used to develop novel THz scanning and imaging devices. 

\section*{Results}
Figure \ref{fig:metasurface} shows the proposed metamirror schematically. A coding metasurface composed of graphene-loaded unit cells is programmed through an FPGA to focus the reflected wave in an arbitrary position. Next, we show the details concerning the design of the unit cell, the coding metasurface, and an imaging device based on the systematic reprogramming of the coding metasurface. 

\subsection*{Unit Cell}
A reflective metasurface operating as a mirror calls for the use of unit cells with a high reflection amplitude and a wide phase range. Fabry-P\'erot resonant structures composed of metallic patches over a square grounded substrate can provide such functionality by adjusting the size or position of the metallic patch to achieve the required phase \cite{Yu2011a}. This approach, however, offers no reconfigurability. In this work, instead, we achieve reconfigurability through changes in the electrostatic bias applied to uniformly sized graphene patches. This modifies the complex conductivity of the graphene sheet and, by extension, the phase of the reflected wave.

\begin{figure}[!ht] 
\centering
\includegraphics[width=1\columnwidth]{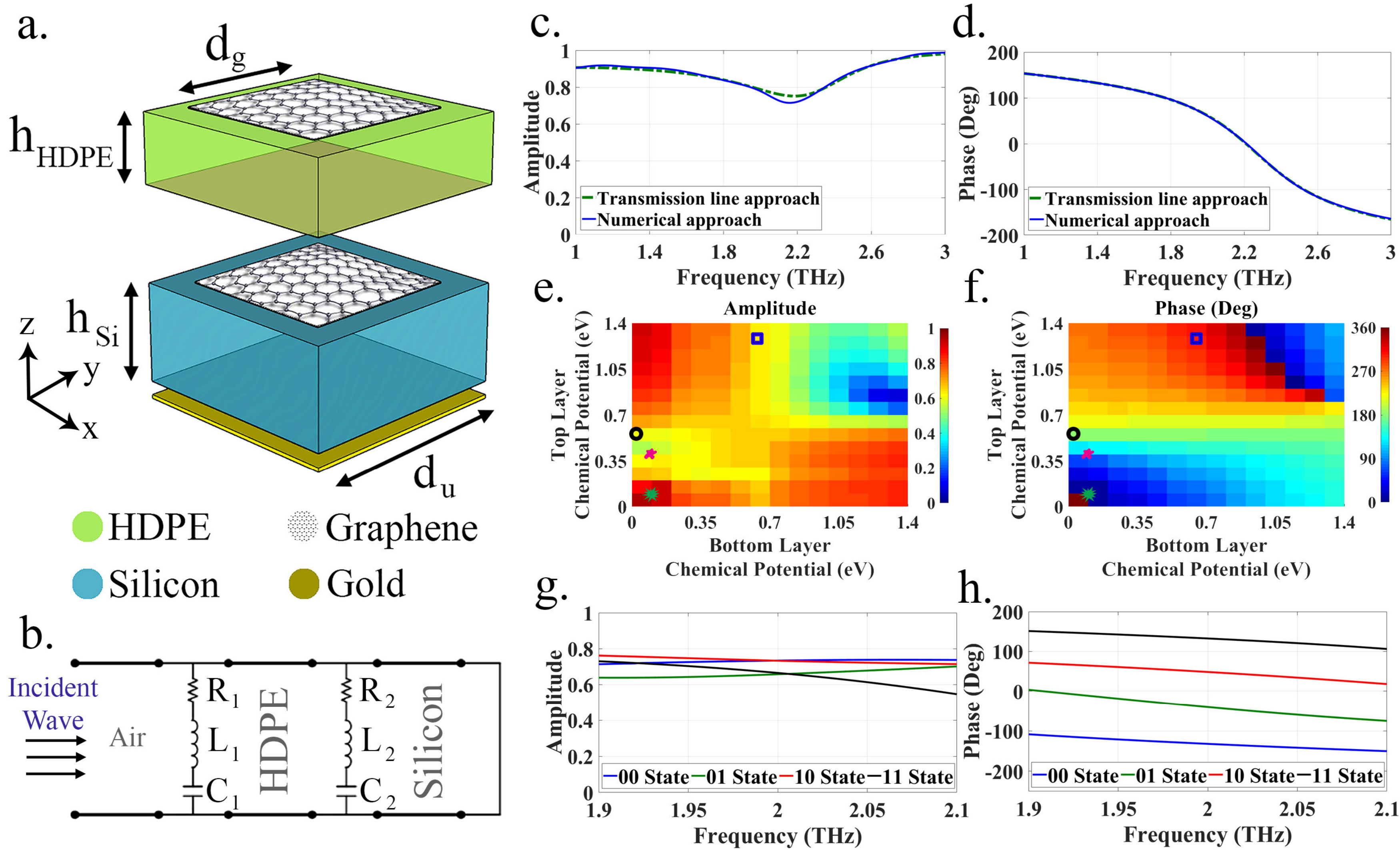} 
\vspace{-0.1cm}
\caption{\textbf{(a)} Dual layer graphene-based unit cell used as building block of the flat metasurface. \textbf{(b)} Approximated circuit model of the unit cell. \textbf{(c, d)} Frequency response of the amplitude and phase of the reflection coefficient of the unit cell composed of squared graphene patches ($d_{g}=16\,\upmu$m) with top and bottom chemical potentials of 0.2 and 0.1 eV, respectively. \textbf{(e, f)} Amplitude and phase of the unit cell response for different top and bottom chemical potentials. \textbf{(g, h)} Resulting unit cell amplitude and phase ($d_{g}=12\,\upmu$m) for the four chosen coding states (which are marked in Fig. \ref{fig:unit cells}(e) and Fig. \ref{fig:unit cells}(f) by different signs): top--bottom chemical potentials are 0.6--0 eV for state 00, 1.3--0.6 eV for state 01, 0.1--0.1 eV for state 10, and 0.4--0.1 for state 11.}
\label{fig:unit cells}
\vspace{-0.5cm}
\end{figure}

Figure \ref{fig:unit cells}(a) shows the proposed unit cell. It maintains the Fabry-P\'erot cavity structure with a graphene sheet over a grounded silicon substrate, yet adding a second graphene sheet over a high-density polyethylene (HDPE) layer. HDPE is chosen due to its particularly low loss behavior in THz band. The unit cells and the graphene patches are dimensioned to provide proper performance at the operation frequency, i.e. 2 THz in this work. The approximated circuit model shown in Fig. \ref{fig:unit cells}(b) is developed in order to provide a theoretical validation of the unit cell behavior. As observed in Figs. \ref{fig:unit cells}(c, d), the theoretical and simulation results show a close agreement. More details on the characteristics of the unit cell and its validation are provided in the \emph{Methods} section.

Compared to single-layer graphene structures, our proposed unit cell based on dual-layer graphene enjoys more flexibility to achieve a wide range of phase response and a high reflection amplitude, simultaneously. The dual layer structure, modeled as two parallel RLC cells, provides an additional degree of freedom as each graphene patch can be biased independently. Fig. \ref{fig:unit cells}(e) and Fig. \ref{fig:unit cells}(f) shows the reflection amplitude and phase response of the unit cell at 2 THz, respectively, as functions of chemical potential of the top and bottom graphene layers ($\mu_{c,1}$ and $\mu_{c,2}$, respectively). A quick exploration reveals a complete $2\pi$ phase range as well as large design space areas with high reflectivity. 

The coding set is built by picking points with as high amplitude as possible, and phases equal to multiples of $2\pi/2^{N_{bit}}$ where $N_{bit}$ is the number of bits. We will see that two bits are enough for our purpose. We choose $\mu_{c,1} = \{0.6, 1.3, 0.1, 0.4\}$ eV and $\mu_{c,2} = \{0, 0.6, 0.1, 0.1\}$ eV corresponding to the bit combinations $B = \{00, 01, 10, 11\}$, respectively. As observed in Fig. \ref{fig:unit cells}(g) and Fig. \ref{fig:unit cells}(h), a phase shift of $\pi/2$ is maintained over the range of 1.9--2.1 THz with an amplitude close to 0.7. This high reflection amplitude for all states guarantees a high focusing efficiency for the designed metamirror.

\subsection*{Coding Metasurface Reflector in the THz Range}
Consider a reflective metasurface under illumination of an incident plane wave at elevation angle $\theta_{i}$ and azimuth angle $\varphi_{i}$. It is used to focus the energy in a given point defined by $\mathbf{r_f}=(x_{f}, y_{f}, z_{f}) \equiv (R_{f}, \theta_{f}, \varphi_{f})$. The coordinate origin is at the geometric center of the metasurface. The generalized Snell's law of reflection is used to formulate the phase profile over the metamirror. The required phase profile $\Phi(x',y')$ at an arbitrary point on the metamirror can be written as
\begin{equation}
\label{eq:rf1}
\Phi(x',y') = \frac{2\pi}{\lambda_0}\mid\mathbf{r'}-\mathbf{r_f}\mid - \mathbf{k_0}\cdot \mathbf{r'},
\end{equation}   
where $\lambda_0$ is the wavelength in free space, $\mathbf{k_0}$ is the incident wavevector, and $\mathbf{r'}=(x', y')$ is the vector position of an arbitrary point on the metasurface.

When the metamirror is illuminated by a normally incident wave ($\theta_{i} = \phi_{i} = 0$), Equation \ref{eq:rf1} leads to
\begin{equation}
\label{eq:rf}
\Phi(x',y') = 
\frac{{2\pi }}{\lambda_{0} }\left( 
\sqrt {{{({R_f}\cos {\theta _f})}^2} + {{(x' - {R_f}\sin {\theta _f}\cos {\varphi _f})}^2} + {{(y' - {R_f}\sin {\theta _f}\sin {\varphi _f})}^2}} 
 - {R_f}\cos {\theta _f}
 \right).
\end{equation}
The actual phase profile of the coding metasurface is obtained by spatially discretizing $\Phi(x',y')$, and rounding off the phase values to those offered by the closest coding states.

\begin{figure}[!ht] 
\centering
\includegraphics[width=1\columnwidth]{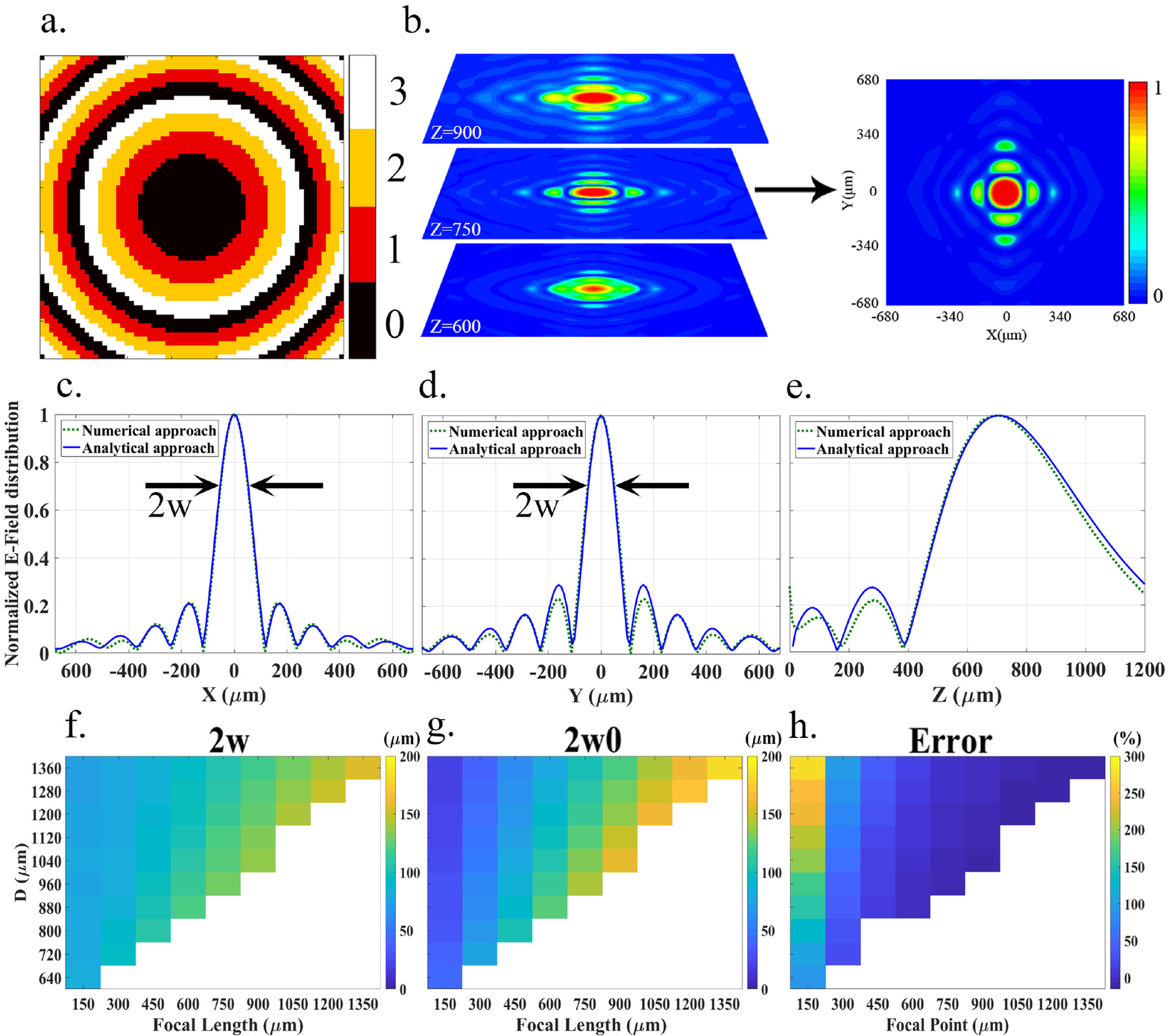} 
\vspace{-0.1cm}
\caption{\textbf{(a)} Arrangement used to achieve the focal point at $x = 0$, $y = 0$, $z = 750\,\upmu$m. \textbf{(b)} Field distribution at different z distances, including the targeted focal depth. \textbf{(c, d, e)} Field distribution along the main radial ($z = 750\,\upmu$m) and axial ($x = y = 0$) directions. \textbf{(f, g, h)} Beam waist $2w$, diffraction limit $2w_{0}$, and relative difference between the waist and diffraction limit, respectively, as functions of the focal length $z$ and lateral size of the metasurface $D$.}
\label{fig:z750}
\vspace{-0.5cm}
\end{figure}

Figure \ref{fig:z750}(a) shows a coding metasurface programmed to focus the beam at $x = 0$, $y = 0$, and $z = 750\,\upmu$m with four states. The size of the metasurface is 68$\times$68 unit cells or 9$\lambda\times$9$\lambda$ at 2 THz. As observed in Fig. \ref{fig:z750}(b), the metasurface achieves effective focusing of the electrical field very close to the desired spot. The actual point of maximum field is (0, 0, 705), which implies an error of around 6\%. By moving away from the focal point, the field spreads out as expected. As demonstrated in Figures \ref{fig:z750}(c--e), side lobes appear around the target focal spot due to the finite and discrete nature of the metasurface. Both the focal point accuracy and the side lobe amplitude can be improved with larger metasurfaces and higher resolution in spatial phase profile. Note that Figures \ref{fig:z750}(c--e) demonstrate close agreement between the numerical results and the theoretical approach, which is detailed in the \emph{Methods} section. 

Figures \ref{fig:z750}(f--h) provide an assessment of the beam waist as a function of the focal length, and the aperture size of the metasurface $D$. To give an instance, the mentioned metasurface with focal length of $750\,\upmu$m and aperture size of $1360\,\upmu$m, the waist diameter of the half-maximum energy density is $2w = 104.9\,\upmu$m, whereas the diffraction-limited full beam waist is given by $2w_{0} = 1.22\lambda_{0}R_{f}/D = 100.9\,\upmu$m. It is worth noting that the proposed metamirror operates close to the diffraction limit even for long focal length, which is generally difficult to achieve. Moreover, larger metasurfaces are apparently required to achieve a larger focal point range, because of the necessity to operate below the Fraunhofer distance $d_{F} = 2D^{2}/\lambda_{0}$.

\begin{figure}[!ht] 
\centering
\includegraphics[width=1\columnwidth]{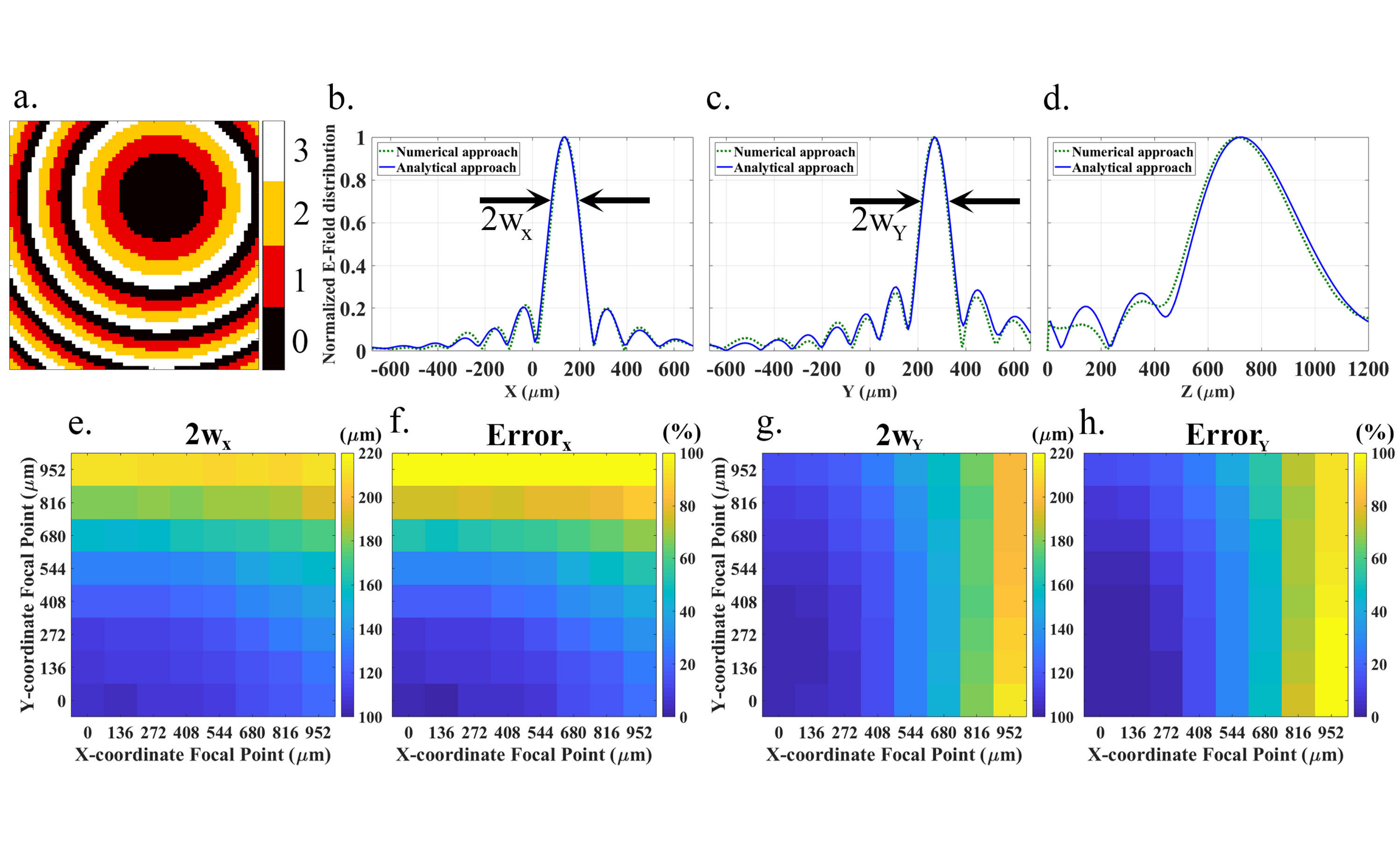} 
\vspace{-0.1cm}
\caption{\textbf{(a)} Coding arrangement used to achieve the focal point at $x = 136\,\upmu$m, $y = 272\,\upmu$m, $z = 750\,\upmu$m. \textbf{(b, c, d)} Field distribution along the main radial ($z = 750\,\upmu$m) and axial ($x = 136\,\upmu$m, $y = 272\,\upmu$m) directions. \textbf{(e, f)} Beam waist in the X direction and relative difference with respect to the diffraction limit, respectively, as functions of the focal point in the $XY$ plane with $z = 750\,\upmu$m. \textbf{(g, h)} Beam waist in the Y direction and relative difference with respect to the diffraction limit, respectively, as functions of the focal point in the $XY$ plane with $z = 750\,\upmu$m.}
\label{fig:xy}
\vspace{-0.5cm}
\end{figure}

Next, Figure \ref{fig:xy}(a) shows how the 68$\times$68 coding metasurface would be programmed to laterally shift the focus to $x = 136\,\upmu$m, $y = 272\,\upmu$m, $z = 750\,\upmu$m. Figures \ref{fig:xy}(b--d) confirm that energy is focused around the desired target with an excellent agreement with theory. The actual position of maximum field is less than 30 $\upmu$m away from the target focal point. Maintaining the focal depth at $z = 750\,\upmu$m and with a lateral size of $D = 1.36$ mm, we perform a beam waist analysis analogous to that of Figure \ref{fig:z750} but in the lateral dimensions. Charts (e--f) from Figure \ref{fig:xy} show the beam waist in the X direction and its relative difference to the diffraction limit as a function of the positions of the focal point, respectively. It is observed that as the focal point is moved away from the metasurface center, especially in the Y direction, the beam waist in the X direction tends to increase since. A complementary behavior is observed for the beam waist in Y direction, as shown in charts (g--h) of Figure \ref{fig:xy}. The reason for such behavior is that the achieved asymmetric phase profile of the metasurface with specified dimensions results in an elliptical beam instead of a circular beam. As a matter of fact, a metasurface with finite dimensions is not capable to produce the necessary complete phase profile to achieve a circular beam pointed away from the metasurface center. As expected, one can operate close to the diffraction limit once being close to the metasurface center.

To note, the electric field around focal point decreases when the Z distance of focal point increase and also when the focal point position moves away from the Z axis which results in a decrease in the focusing efficiency.

\subsection*{Programmable Metamirror for 3D Confocal Terahertz Imaging}
The results obtained above demonstrate that the proposed metamirror can set the focus to any point in a distance from the metamirror. Due to the unequivocal relation between the phase profile and focal point given in \eqref{eq:rf}, it is relatively easy to develop algorithms that scan the focal point of the metamirror to perform three-dimensional confocal imaging. 

\begin{figure}[!ht] 
\centering 
\includegraphics[width=1\columnwidth]{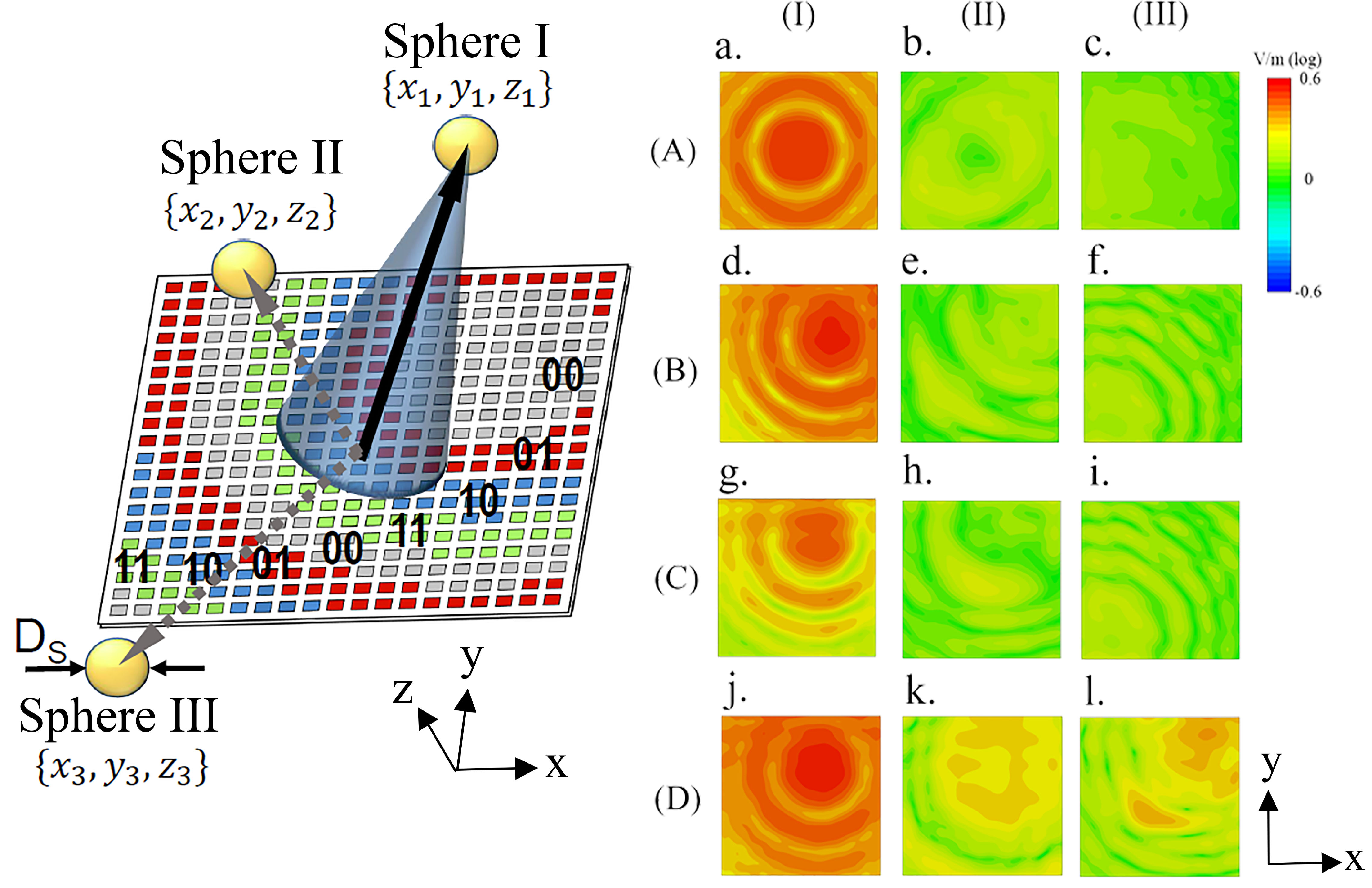} 
\vspace{-0.1cm}
\caption{Coding metasurface programmed to focus on the Sphere I. Spheres II and III are assumed to be located relatively close (near focus) and far (outside focus) from Sphere I, respectively. The plots represent the field distributions at the far-field of the metasurface, whereas the plots in the different rows consider distinct locations (A, B, C, D) of the spheres. The locations are $\{0, 0, 750\}$, $\{200, 200, 750\}$ and $\{-500, -500, 750\}$ for case A; $\{272, 272, 750\}$, $\{500, 500, 750\}$ and $\{-500, -500, 750\}$ for case B; $\{136, 408, 750\}$, $\{300, 550, 750\}$ and $\{-500, -500, 750\}$ for case C; and $\{136, 272, 750\}$, $\{136, 272, 300\}$ and $\{300, 500, 1100\}$ for case D. All dimensions are in $\upmu$m units. The same color bar is used for all cases.}
\label{fig:imaging}
\vspace{-0.5cm}
\end{figure}

Figure \ref{fig:imaging} exemplifies the confocal imaging concept by the proposed metamirror. Let us consider three spherical objects, as point scatterers, placed within the area of the metasurface influence. Let us also assume that the diameter of the spheres $D_{S}$ is commensurate to the beam waist. By means of the principle of reversibility of the optical path, a strong reflected field should be observed once a plane wave is focused onto a sphere by the metasurface. On the contrary, weak reflected field should be observed when the sphere is absent on the focal point. Thus, one can make a 3D map (image) of the reflected field response when the focal point is scanned in space. Differences in the reflected field response due to the scan of focal point creates the confocal image.

Charts (a--l) of Figure \ref{fig:imaging} show the reflected field distributions at the far-field of the metasurface with assuming that sphere I is at the focal point (first column), sphere II is close to the focal point (second column), and sphere III is far from the focal point (third column). Each row assumes different locations for all the spheres. From different plots, it is clear that strong and weak reflected field distribution can be observed when the spheres are present and absent at the focal point, respectively. It confirms that the confocal imaging scheme works well in a variety of cases: (A) focus in the center of the metasurface, (B) focus not centered, (C) focus not centered and sphere II closer, and (D) focus not centered and sphere II in a different depth. 

\section*{Discussion}
In this work, we have presented for the first time the design of a fully reprogrammable metamirror that can be used in terahertz confocal imaging. The metamirror acts as a focusing reflectarray that leverages the tunability of graphene to achieve the required phase range and programmability at the unit cell level. At the metasurface level, we adopted a coding metasurface approach that allows to dynamically shift the focal point by adjusting the center and slope of a concentric phase gradient. This is in stark contrast to the design by Huang \emph{et al.}\cite{Huang2017a}, wherein a single graphene layer is laid over the whole metasurface area and tuned using a single bias source. Such approach only allows to control the focal depth with limited range. Focusing with programmable coding metasurfaces has been only demonstrated in the GHz range with devices that use PIN diodes for reconfiguration. This limits the amount of achievable unit cell states, reducing the precision and focusing range.

We demonstrated analytically and numerically that the proposed metamirror can shift the focal point both laterally and in depth. With 2 bits per unit cell and assuming a total size of $9\lambda \times 9\lambda$, energy is focused around the desired focal point with less than 6\% deviation in the evaluated points. Moreover, the metamirror can operate close to the diffraction limit as long as the focal point is located around the focal depth and centered with respect to the metasurface. We also showed that the programmability of the focal point can be applied to the development of compact devices for 3D terahertz confocal imaging. 

As a final note, it is worth highlighting that this work brings the concept close to realization by making reasonable assumptions. First, we consider a graphene quality leading to a relaxation time of $\tau = 0.6$ ps, which is well achievable with Chemical Vapor Deposition (CVD) techniques and encapsulated in hexagonal boron nitride (h-BN)\cite{Banszerus2015, Schmitz2017}. Those techniques in fact produce significantly higher relaxation times, which would lead to lower loss and better overall performance. Moreover, considering the electrically thin boron nitride layer has a negligible impact in the simulation results.To mention, when the graphene has a lower relaxation time, the plasmonic losses within the structure increase. Therefore, it results in a device with a lower efficiency. Regarding the biasing scheme, different gate-controllable designs have been simulated\cite{Huang2012ARRAY} and experimentally validated\cite{Gomez2015}, which could be adapted to local addressing of unit cells. . Before actually addressing fabrication, a challenge to be addressed in future work would consist on minimizing the interference imposed by the metallic biasing structure on the response of the metasurface. It is worthy to note that bilayer graphene, a material consisting of two layers of graphene, exists in the various forms such as Bernal-stacked form \cite{yan2011-}. Depending on the relative position of layers, different quantum effects can happen in the spacing. But what we have here is different. In our proposed dual-layer structure, the insulating layer (here HDPE) between two graphene patches is sufficiently large for quantum effects to be negligible \cite{tamagnone2012reconfigurable}. Therefore, crystallographic alignment is not required, and micrometric precision of current lithographic techniques is enough to achieve the results reported in the paper. In summary, the process of fabrication of dual-layer metasurface structure is generally similar to one-layer structure; and the existing transfer printing methods could be employed \cite{Lee2016, Gomez2015}.

\section*{Methods}
\subsection*{Unit cell modeling}
The unit cell structure shown in in Figure \ref{fig:unit cells}(a) is implemented in CST Microwave Studio \cite{CST}. The length of the unit cell is $d_{u}=20\,\upmu$m and graphene patches are squared with side $d_{g}=12\,\upmu$m. The operation frequency is set to 2 THz. We consider bulk silicon as the substrate with relative permittivity $\varepsilon_{r}=11.9$, loss tangent $\tan(\delta)=2.5\times 10^{-4}$, and thickness $h_{Si}=10\,\upmu$m. The extra layer is composed of a layer of HDPE with relative permittivity $\varepsilon_{r}=2.37$, loss tangent $\tan(\delta)=0.011$, and thickness $h_{HDPE}=4\,\upmu$m.

The graphene layers are modeled as infinitesimally thin sheets with surface impedance $Z = 1/\sigma(\omega)$, where $\sigma(\omega)$ is the frequency-dependent complex conductivity of graphene. The conductivity is given by 
\begin{equation}
\sigma\left(\omega\right)=\frac{2e^{2}}{\pi\hbar}\frac{k_{B}T}{\hbar}\ln\left[2\cosh\left[\frac{\mu_{c}}{2k_{B}T}\right]\right]\frac{i}{\omega+i\tau^{-1}},\label{eq:sigma_graphene}
\end{equation}
where $e$, $\hbar$ and $k_{B}$ are constants corresponding to the charge of an electron, the reduced Planck constant and the Boltzmann constant, respectively \cite{Hanson2008}. Variables $T$, $\tau$ and $\mu_{c}$ correspond to the temperature ($T = 300$ K in this paper), the relaxation time and the chemical potential of the graphene layer. Note that this expression neglects effects at the graphene edges and considers that the Drude-like intraband contribution dominates, which are experimentally validated assumptions at the sizes and frequencies considered in this work \cite{AbadalTCOM}. In all cases, the relaxation time of graphene $\tau = 0.6$ ps, which is well achievable with current fabrication and encapsulation techniques \cite{Banszerus2015}. The chemical potential of the top and bottom layers are initially left as parameters for exploration and then fixed to discrete values leading to the adequate amplitude and phase. 

\subsection*{Simulation methods}
We implement the unit cells and the complete metasurface in CST Microwave Studio \cite{CST}. For the simulation of unit cells, a Floquet port is applied to excite an x-polarized normal plane wave onto the unit cell and the back-scattered wave is then measured. This way, we obtain the amplitude and phase of the reflection coefficient of the proposed unit cells while taking into consideration mutual coupling between adjacent unit cells. After obtaining the response of individual unit cells, the complete metamirror is modeled as an array of such unit cells with different phases. The metasurface is illuminated by a normal plane wave with x-polarization to obtain the scattered fields above the metasurface and, thus, the field distribution around the focal point. Finally, we simulate the THz imaging device by estimating the scattering response of the spheres in different positions. To this end, we again consider x-polarized normal illumination. To identify the presence of a sphere in the focal point of the metamirror, we compare the far field response of the metamirror with and without the spherical objects. Strong variations indicate the presence of a sphere in the focal point of the metamirror. To note, for the results shown in Figures \ref{fig:z750} and \ref{fig:xy}, the reflected fields i.e. the difference between the total wave and incident wave are extracted to survey the achieved focal points while for the results shown in Figure \ref{fig:imaging}, the difference between the reflected fields with and without the sphere are obtained to identify the presence of the spherical object in the focal point of the metamirror.  

\subsection*{Theoretical validation}
To gain physical insight into the scattering behavior of the proposed unit cell, an approximate circuit model has been developed. As sketched in Figure \ref{fig:unit cells}(b), the graphene patches are modeled as RLC series circuits ($R_{1}$, $L_{1}$, $C_{1}$; $R_{2}$, $L_{2}$, $C_{2}$) in parallel with equivalent transmission lines representing the HDPE and grounded silicon layers. The values of
resistance, inductance, and capacitance of the graphene patches have been extracted from the reflection parameter at the boundary between air and unit cell, which only depend on the chemical potentials of the top and bottom graphene layers for a fixed patch size. For the results shown in Figure \ref{fig:unit cells}(c), (d), ($R_{1}$, $L_{1}$, $C_{1}$; $R_{2}$, $L_{2}$, $C_{2}$) are (20.32 \textOmega, 0.013 pH, 0.107 fF; 35.97 \textOmega, 26.26 pH, 3.38 fF). 

To validate the response of the metamirror around the focal point, we need to use analytical models compatible with near field conditions. Fresnel diffraction theory cannot be used because of the small focal length to lens diameter ratio of our device. Alternatively, we apply the Huygens' principle to model the currents excited at the metasurface as electrically small sources of radiation. Consider an incident wave polarized along the x-direction that excites an x-directed current density $J_{x}$ on all unit cells. Therefore, we can write the inhomogeneous potential wave equation as
\begin{equation}
\label{eq:theory1}
\nabla^{2}A_{x} + k^{2}A_{x} = {-\mu_{0}} J_{x}(x', y').
\end{equation}
Since the unit cells are electrically small, it is reasonable to assume each unit cell as a punctual source. Consequently, the Green's function for the wave equation allows us to write the solution of Equation \eqref{eq:theory1} as
\begin{equation}
\label{eq:theory2}
A_{x} = \frac{\mu_{0}}{4\pi} J_{x}(x', y')\frac{e^{-jk\mid \mathbf{r}-\mathbf{r'}\mid}}{\mid \mathbf{r}-\mathbf{r'}\mid}.
\end{equation}
where $\mathbf{r}=(x,y,z)$ and $\mathbf{r'}=(x',y')$ are the vector positions of the observation and source points, respectively. The magnetic and electric fields can be found by applying $ H = \tfrac{1}{\mu_{0}}\nabla\times A$ and $E = \tfrac{1}{j\omega \varepsilon_{0}}\nabla\times H$, respectively. This allows to obtain a closed form expression for the electrical field
\begin{equation}
\begin{split}
&E_{x}(x,y,z) = \frac{-j\eta}{4\pi k}(G_{1} + (x-x')^{2}G_{2}) J_{x} e^{-jk\mid \mathbf{r}-\mathbf{r'}\mid} \\
&G_{1} = \frac{-1-jk\mid \mathbf{r}-\mathbf{r'}\mid+k^2 \mid \mathbf{r}-\mathbf{r'}\mid^2}{\mid \mathbf{r}-\mathbf{r'}\mid^3} \\
&G_{2} = \frac{3+j3k\mid \mathbf{r}-\mathbf{r'}\mid-k^2\mid \mathbf{r}-\mathbf{r'}\mid^2}{\mid \mathbf{r}-\mathbf{r'}\mid^5}
\end{split}
\end{equation}
Finally, the metasurface response is obtained by adding the contribution of its $M\times N$ cells as
\begin{equation}
\begin{split}
&E_{x}(x,y,z) = \frac{-j\eta}{4\pi k}\sum_{m=1}^{M}\sum_{n=1}^{N} (G_{1} + (x-x')^2 G_{2}) \exp[+j(\Phi(m,n)-k\mid \mathbf{r}-\mathbf{r'}\mid)] \\
&\mid \mathbf{r}-\mathbf{r'}\mid = \sqrt{(x-D(m-\tfrac{1}{2}))^2 + (y-D(n-\tfrac{1}{2}))^2 + z^2}
\end{split}
\end{equation}

\section*{Acknowledgements}
This work has been partially funded by Iran's National Elites Foundation (INEF), the Iran National Science Foundation (INSF) Chair of Computational Electromagnetics and Bio-Electromagnetics, the Spanish Ministry of \emph{Econom\'ia y Competitividad} under grant PCIN-2015-012, and by the Catalan Institution for Research and Advanced Studies (ICREA), and by the European Union via the Horizon 2020: Future Emerging Topics call (FET Open), grant EU736876, project VISORSURF.

\end{document}